\pdfoutput=1

\documentclass[12pt,a4paper]{article}

\usepackage{ifthen} 
\newboolean{pdflatex}
\setboolean{pdflatex}{true} 

\newboolean{articletitles}
\setboolean{articletitles}{true} 

\newboolean{uprightparticles}
\setboolean{uprightparticles}{false} 

\newboolean{inbibliography}
\setboolean{inbibliography}{false} 

\usepackage[top=1in, bottom=1.25in, left=1in, right=1in]{geometry}

\usepackage{microtype}
\usepackage{lineno}  
\usepackage{xspace} 
\usepackage{caption} 

\usepackage{graphicx}  
\usepackage{color}
\usepackage{colortbl}
\graphicspath{{./figs/}} 

\usepackage{amsmath} 
\usepackage{amssymb}
\usepackage{amsfonts}
\usepackage{upgreek} 

\newcommand*\patchAmsMathEnvironmentForLineno[1]{%
\expandafter\let\csname old#1\expandafter\endcsname\csname #1\endcsname
\expandafter\let\csname oldend#1\expandafter\endcsname\csname
end#1\endcsname
 \renewenvironment{#1}%
   {\linenomath\csname old#1\endcsname}%
   {\csname oldend#1\endcsname\endlinenomath}%
}
\newcommand*\patchBothAmsMathEnvironmentsForLineno[1]{%
  \patchAmsMathEnvironmentForLineno{#1}%
  \patchAmsMathEnvironmentForLineno{#1*}%
}
\AtBeginDocument{%
\patchBothAmsMathEnvironmentsForLineno{equation}%
\patchBothAmsMathEnvironmentsForLineno{align}%
\patchBothAmsMathEnvironmentsForLineno{flalign}%
\patchBothAmsMathEnvironmentsForLineno{alignat}%
\patchBothAmsMathEnvironmentsForLineno{gather}%
\patchBothAmsMathEnvironmentsForLineno{multline}%
\patchBothAmsMathEnvironmentsForLineno{eqnarray}%
}

\usepackage{hyperref}    
\usepackage[all]{hypcap} 

\usepackage{xspace} 
\usepackage{upgreek}

\def\lhcb {\mbox{LHCb}\xspace}

\def\babar  {\mbox{BaBar}\xspace}
\def\belle  {\mbox{Belle}\xspace}
\def\cleo   {\mbox{CLEO}\xspace}

\def\aleph  {\mbox{ALEPH}\xspace}

\def\MagUp {\mbox{\em Mag\kern -0.05em Up}\xspace}

\ifthenelse{\boolean{uprightparticles}}%
{

 \def\Pnu         {\ensuremath{\upnu}\xspace}                 
                  
 \def\Ppi         {\ensuremath{\uppi}\xspace}

 \def\PDelta      {\ensuremath{\Delta}\xspace}                 
 \def\PXi      {\ensuremath{\Xi}\xspace}                 
 \def\PLambda      {\ensuremath{\Lambda}\xspace}                 
 \def\PSigma      {\ensuremath{\Sigma}\xspace}                 
 \def\POmega      {\ensuremath{\Omega}\xspace}                 
 \def\PUpsilon      {\ensuremath{\Upsilon}\xspace}                 
 
 \def\PB      {\ensuremath{\mathrm{B}}\xspace}                 
                  
 \def\PD      {\ensuremath{\mathrm{D}}\xspace}

 \def\PK      {\ensuremath{\mathrm{K}}\xspace}

 \def\Pb      {\ensuremath{\mathrm{b}}\xspace}                 
 \def\Pc      {\ensuremath{\mathrm{c}}\xspace}

 \def\Pi      {\ensuremath{\mathrm{i}}\xspace}

 \def\Pp      {\ensuremath{\mathrm{p}}\xspace}

 \def\Ps      {\ensuremath{\mathrm{s}}\xspace}

}
{

 \def\Pnu         {\ensuremath{\nu}\xspace}                 
                  
 \def\Ppi         {\ensuremath{\pi}\xspace}

 \mathchardef\PDelta="7101
 \mathchardef\PXi="7104
 \mathchardef\PLambda="7103
 \mathchardef\PSigma="7106
 \mathchardef\POmega="710A
 \mathchardef\PUpsilon="7107
                  
 \def\PB      {\ensuremath{B}\xspace}                 
                  
 \def\PD      {\ensuremath{D}\xspace}

 \def\PK      {\ensuremath{K}\xspace}

 \def\Pb      {\ensuremath{b}\xspace}                 
 \def\Pc      {\ensuremath{c}\xspace}

 \def\Pi      {\ensuremath{i}\xspace}

 \def\Pp      {\ensuremath{p}\xspace}

 \def\Ps      {\ensuremath{s}\xspace}

}

\makeatletter
\ifcase \@ptsize \relax
  \newcommand{\miniscule}{\@setfontsize\miniscule{4}{5}}
\or
  \newcommand{\miniscule}{\@setfontsize\miniscule{5}{6}}
\or
  \newcommand{\miniscule}{\@setfontsize\miniscule{5}{6}}
\fi
\makeatother

\DeclareRobustCommand{\optbar}[1]{\shortstack{{\miniscule (\rule[.5ex]{1.25em}{.18mm})}
  \\ [-.7ex] $#1$}}

\def\ellp       {{\ensuremath{\ell^+}}\xspace}

\def\neu        {{\ensuremath{\Pnu}}\xspace}
\def\neub       {{\ensuremath{\overline{\Pnu}}}\xspace}
\def\neul       {{\ensuremath{\neu_\ell}}\xspace}
\def\neulb      {{\ensuremath{\neub_\ell}}\xspace}

\def\squark    {{\ensuremath{\Ps}}\xspace}

\def\cquark    {{\ensuremath{\Pc}}\xspace}

\def\bquark    {{\ensuremath{\Pb}}\xspace}

\def\pion   {{\ensuremath{\Ppi}}\xspace}

\def\pip    {{\ensuremath{\pion^+}}\xspace}
\def\pim    {{\ensuremath{\pion^-}}\xspace}

\def\pimp   {{\ensuremath{\pion^\mp}}\xspace}

\def\kaon    {{\ensuremath{\PK}}\xspace}
  \def\Kbar    {{\kern 0.2em\overline{\kern -0.2em \PK}{}}\xspace}

\def\KorKbar    {\kern 0.18em\optbar{\kern -0.18em K}{}\xspace}

\def\Kp      {{\ensuremath{\kaon^+}}\xspace}
\def\Km      {{\ensuremath{\kaon^-}}\xspace}
\def\Kpm     {{\ensuremath{\kaon^\pm}}\xspace}

  \def\Dbar    {{\kern 0.2em\overline{\kern -0.2em \PD}{}}\xspace}
\def\D       {{\ensuremath{\PD}}\xspace}

\def\DorDbar    {\kern 0.18em\optbar{\kern -0.18em D}{}\xspace}
\def\Dz      {{\ensuremath{\D^0}}\xspace}

\def\B       {{\ensuremath{\PB}}\xspace}
\def\Bbar    {{\ensuremath{\kern 0.18em\overline{\kern -0.18em \PB}{}}}\xspace}

\def\BorBbar    {\kern 0.18em\optbar{\kern -0.18em B}{}\xspace}
\def\Bz      {{\ensuremath{\B^0}}\xspace}

\def\Bu      {{\ensuremath{\B^+}}\xspace}

\def\Bd      {{\ensuremath{\B^0}}\xspace}
\def\Bs      {{\ensuremath{\B^0_\squark}}\xspace}

  \def\Y#1S{\ensuremath{\PUpsilon{(#1S)}}\xspace}

\def\proton      {{\ensuremath{\Pp}}\xspace}
\def\antiproton  {{\ensuremath{\overline \proton}}\xspace}

\def\Lz          {{\ensuremath{\PLambda}}\xspace}
\def\Lbar        {{\ensuremath{\kern 0.1em\overline{\kern -0.1em\PLambda}}}\xspace}
\def\LorLbar    {\kern 0.18em\optbar{\kern -0.18em \PLambda}{}\xspace}

\def\Lc      {{\ensuremath{\Lz^+_\cquark}}\xspace}

\def\BF         {{\ensuremath{\mathcal{B}}}\xspace}

\newcommand{\decay}[2]{\ensuremath{#1\!\to #2}\xspace}         

\def\to                 {\ensuremath{\rightarrow}\xspace}

\def\CP                {{\ensuremath{C\!P}}\xspace}

\def\AT#1     {\ensuremath{A_{\mathrm{T}}^{#1}}\xspace}           

\def\C#1      {\ensuremath{\mathcal{C}_{#1}}\xspace}                       
\def\Cp#1     {\ensuremath{\mathcal{C}_{#1}^{'}}\xspace}                    
\def\Ceff#1   {\ensuremath{\mathcal{C}_{#1}^{\mathrm{(eff)}}}\xspace}        
\def\Cpeff#1  {\ensuremath{\mathcal{C}_{#1}^{'\mathrm{(eff)}}}\xspace}       
\def\Ope#1    {\ensuremath{\mathcal{O}_{#1}}\xspace}                       
\def\Opep#1   {\ensuremath{\mathcal{O}_{#1}^{'}}\xspace}                    

\newcommand{\tev}{\ifthenelse{\boolean{inbibliography}}{\ensuremath{~T\kern -0.05em eV}}{\ensuremath{\mathrm{\,Te\kern -0.1em V}}}\xspace}
\newcommand{\gev}{\ensuremath{\mathrm{\,Ge\kern -0.1em V}}\xspace}
\newcommand{\mev}{\ensuremath{\mathrm{\,Me\kern -0.1em V}}\xspace}
\newcommand{\kev}{\ensuremath{\mathrm{\,ke\kern -0.1em V}}\xspace}
\newcommand{\ev}{\ensuremath{\mathrm{\,e\kern -0.1em V}}\xspace}
\newcommand{\gevc}{\ensuremath{{\mathrm{\,Ge\kern -0.1em V\!/}c}}\xspace}
\newcommand{\mevc}{\ensuremath{{\mathrm{\,Me\kern -0.1em V\!/}c}}\xspace}
\newcommand{\gevcc}{\ensuremath{{\mathrm{\,Ge\kern -0.1em V\!/}c^2}}\xspace}
\newcommand{\gevgevcccc}{\ensuremath{{\mathrm{\,Ge\kern -0.1em V^2\!/}c^4}}\xspace}
\newcommand{\mevcc}{\ensuremath{{\mathrm{\,Me\kern -0.1em V\!/}c^2}}\xspace}

\def\mum  {\ensuremath{{\,\upmu\mathrm{m}}}\xspace}

\def\invfb   {\ensuremath{\mbox{\,fb}^{-1}}\xspace}

\newcommand{\chisq}{\ensuremath{\chi^2}\xspace}

\newcommand{\chisqip}{\ensuremath{\chi^2_{\text{IP}}}\xspace}

\def\gsim{{~\raise.15em\hbox{$>$}\kern-.85em
          \lower.35em\hbox{$\sim$}~}\xspace}
\def\lsim{{~\raise.15em\hbox{$<$}\kern-.85em
          \lower.35em\hbox{$\sim$}~}\xspace}

\def\sPlot{\mbox{\em sPlot}\xspace}

\def\ptot       {\mbox{$p$}\xspace}
\def\pt         {\mbox{$p_{\mathrm{ T}}$}\xspace}

\def\tell1  {TELL1\xspace}
\def\ukl1   {UKL1\xspace}

\def\Bds{{\ensuremath{\B^0_{\kern -0.1em{\scriptscriptstyle (}\kern -0.05em\squark\kern -0.03em{\scriptscriptstyle )}}}}\xspace}
\newcommand{\BPPbar}{\texorpdfstring{\decay{\Bds}{\proton \antiproton}}{}}
\newcommand{\BdPPbar}{\texorpdfstring{\decay{\Bd}{\proton \antiproton}}{}}
\newcommand{\BsPPbar}{\texorpdfstring{\decay{\Bs}{\proton \antiproton}}{}}
\newcommand{\BdKPi}{\texorpdfstring{\decay{\Bd}{\Kp \pim}}{}}

\newcommand{\BdPiPi}{\texorpdfstring{\decay{\Bd}{\pip \pim}}{}}

\newcommand{\BsKK}{\texorpdfstring{\decay{\Bs}{\Kp \Km}}{}}

\newcommand{\BsPiK}{\texorpdfstring{\decay{\Bs}{\pip \Km}}{}}
\newcommand{\LbPPi}{\texorpdfstring{\decay{\PLambda_b^0}{\proton \pim}}{}}
\newcommand{\LPPi}{\texorpdfstring{\decay{\PLambda}{\proton \pim}}{}}
\newcommand{\LbPK}{\texorpdfstring{\decay{\PLambda_b^0}{\proton \Km}}{}}

\newcommand{\BuPPbarK}{\texorpdfstring{\decay{\Bu}{\proton \antiproton \Kp}}{}}

\newcommand{\BuPLbar}{\texorpdfstring{\decay{\Bu}{\proton \Lbar}}{}}
\newcommand{\BuPLbarExcited}{\texorpdfstring{\decay{\Bu}{\proton \Lbar(1520)}}{}}

\newcommand{\LcPKPi}{\texorpdfstring{\decay{\Lc}{\proton \Km \pip}}{}}

\newcommand{\BuPPbarLNu}{\texorpdfstring{\decay{\Bu}{\proton \antiproton \ellp \neulb}}{}}

\def\PPbar {$\proton\antiproton$\xspace}
\def\KPi   {$\Kp\pim$\xspace}
\def\BothChargeKPi   {$\Kpm\pimp$\xspace}

\usepackage{cite} 
\usepackage{mciteplus}

\usepackage{longtable} 

\begin{document}

\renewcommand{\thefootnote}{\fnsymbol{footnote}}
\setcounter{footnote}{1}


\begin{titlepage}
\pagenumbering{roman}

\vspace*{-1.5cm}
\centerline{\large EUROPEAN ORGANIZATION FOR NUCLEAR RESEARCH (CERN)}
\vspace*{1.5cm}
\noindent
\begin{tabular*}{\linewidth}{lc@{\extracolsep{\fill}}r@{\extracolsep{0pt}}}
\ifthenelse{\boolean{pdflatex}}
{\vspace*{-2.7cm}\mbox{\!\!\!\includegraphics[width=.14\textwidth]{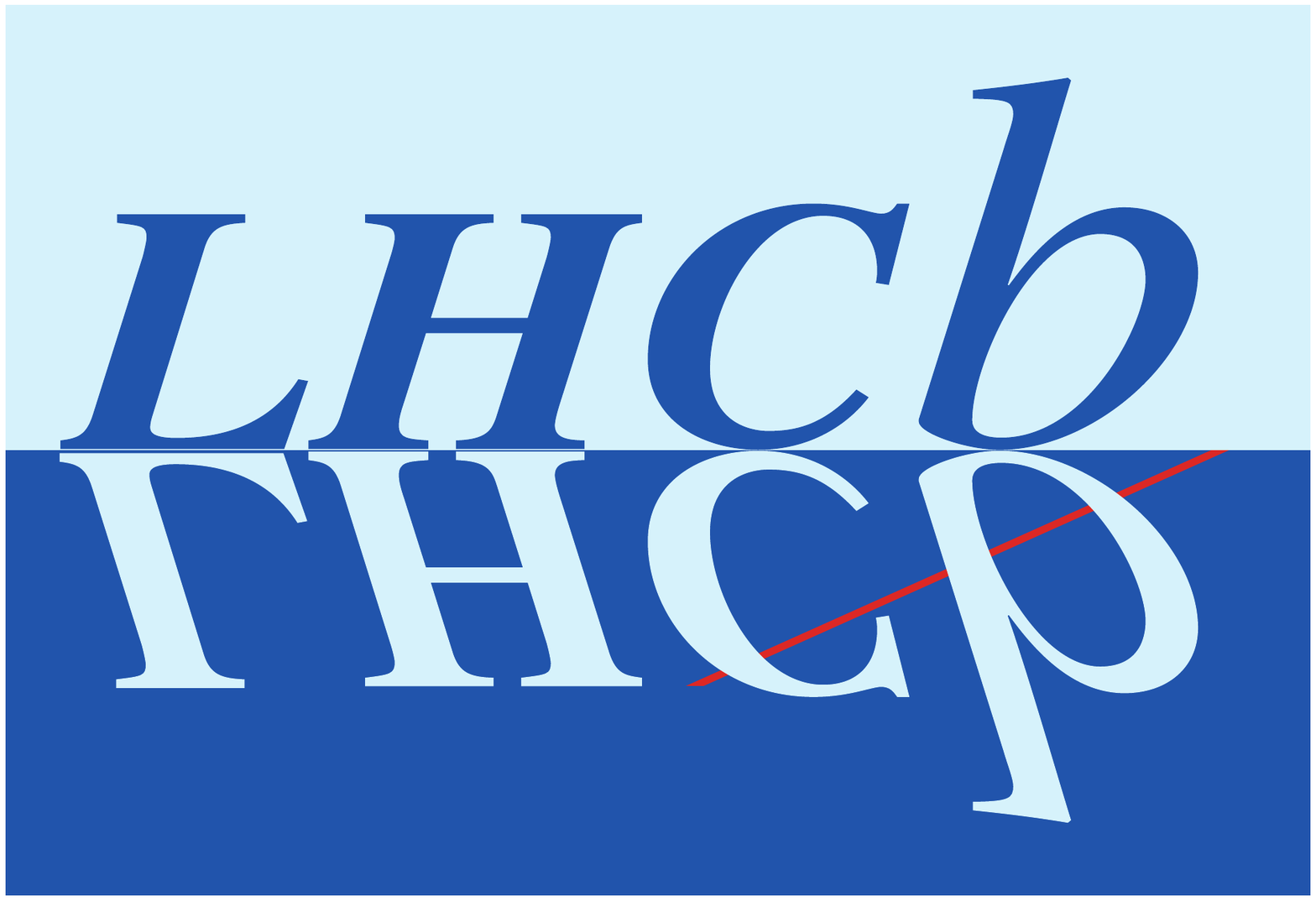}} & &}%
{\vspace*{-1.2cm}\mbox{\!\!\!\includegraphics[width=.12\textwidth]{lhcb-logo.eps}} & &}%
\\
 & & CERN-EP-2017-190 \\  
 & & LHCb-PAPER-2017-022 \\  
 & & 4 December 2017 \\ 
 & & \\
\end{tabular*}

\vspace*{4.0cm}

{\normalfont\bfseries\boldmath\huge
\begin{center}
First observation of the rare\\
purely baryonic decay \BdPPbar
\end{center}
}

\vspace*{2.0cm}

\begin{center}
The LHCb collaboration\footnote{Authors are listed at the end of this Letter.}
\end{center}

\vspace{\fill}

\begin{abstract}
\noindent
The first observation of the decay of a \Bz meson to a purely baryonic
final state, \BdPPbar, is reported.
The proton-proton collision data sample used was collected with the \lhcb experiment at
center-of-mass energies of 7 and 8\tev and corresponds to an integrated luminosity of 3.0\invfb.
The branching fraction is determined to be
$$
\BF(\BdPPbar) = ( 1.25 \pm 0.27 \pm 0.18 ) \times 10^{-8} \,,
$$
where the first uncertainty is statistical and the second systematic.
The decay mode \BdPPbar is the rarest decay of the $B^0$ meson observed to date.
The decay \BsPPbar is also investigated. No signal is seen and the upper limit
$\BF( \BsPPbar) < 1.5 \times 10^{-8}$ at 90\% confidence level
is set on the branching fraction.
\end{abstract}

\vspace*{2.0cm}

\begin{center}
  Published in Phys.~Rev.~Lett. 119 (2017) 232001
\end{center}

\vspace{\fill}

{\footnotesize 
\centerline{\copyright~CERN on behalf of the \lhcb collaboration, licence \href{http://creativecommons.org/licenses/by/4.0/}{CC-BY-4.0}.}}
\vspace*{2mm}

\end{titlepage}


\newpage
\setcounter{page}{2}
\mbox{~}
%
%
\cleardoublepage


\renewcommand{\thefootnote}{\arabic{footnote}}
\setcounter{footnote}{0}



\pagestyle{plain} 
\setcounter{page}{1}
\pagenumbering{arabic}


Studies of $B$ mesons decaying to baryonic final states have been carried out
since the late 1990s~\cite{PhysRevLett.79.3125}.
It was quickly realized that baryonic and mesonic $B$-meson decays differ in a number of ways.
Two-body baryonic decays are suppressed with respect to decays to multibody final states~\cite{Hsiao:2014zza,Cheng:2014qxa}
and the characteristic threshold enhancement in the baryon-antibaryon
mass spectrum~\cite{Hou:2000bz,Bevan:2014iga} is still not fully understood.
The study of such decays
provides information on the dynamics of $B$ decays and tests QCD-based models
of the hadronization process~\cite{Bevan:2014iga}.
It helps to discriminate the available models
and makes it possible to extract both tree and penguin amplitudes of charmless two-body baryonic
decays when combining the information on the \BdPPbar and \BuPLbar branching fractions~\cite{Chua:2016aqy}.

Baryonic $B$ decays are also interesting in the study of \CP violation.
First evidence of \CP violation in baryonic $B$ decays has been reported
from the analysis of \BuPPbarK decays~\cite{LHCb-PAPER-2014-034}
and awaits confirmation in other decay modes~\cite{Geng:2006jt}.

This Letter presents a search for the suppressed decays of \Bz and \Bs mesons
to the two-body charmless baryonic final state $\proton \antiproton$.
Prior to searches at the LHC, the \aleph, \cleo, \babar and \belle collaborations
searched for the \BdPPbar decay~\cite{Buskulic1996471,Coan:1998hy,Aubert:2004fy,Tsai:2007pp}.
The most stringent upper limit on its branching fraction
was obtained by the \belle experiment and is
$\BF( \BdPPbar) < 1.1 \times 10^{-7}$ at 90$\%$ confidence level (CL)~\cite{Tsai:2007pp}.
The only search for the \BsPPbar decay,
performed by the \aleph collaboration,
yielded the upper limit $\BF( \BsPPbar) < 5.9 \times 10^{-5}$
at 90\% CL~\cite{Buskulic1996471}.

The \lhcb collaboration has greatly increased the knowledge of baryonic
$B$ decays in recent years~\cite{LHCb-PAPER-2013-038,LHCb-PAPER-2014-034,LHCb-PAPER-2014-039,LHCb-PAPER-2016-048,LHCb-PAPER-2017-012,LHCb-PAPER-2017-005}.
The collaboration has reported the first observation of a two-body charmless
baryonic \Bu decay, \BuPLbarExcited~\cite{LHCb-PAPER-2014-034},
and the first evidence for \BdPPbar, a two-body charmless baryonic decay of the \Bd meson~\cite{LHCb-PAPER-2013-038}.
The experimental data on two-body final states is nevertheless scarce.
The study of these suppressed modes requires large data samples that are
presently only available at the LHC.

In this analysis, in order to suppress common systematic uncertainties,
the branching fractions of the \BdPPbar and \BsPPbar decays are measured using
the topologically identical decay \BdKPi.
The branching fractions are determined from
\begin{equation}
\BF(\BPPbar) = \frac{N(\BPPbar)}{N(\BdKPi)}\,
               \frac{\varepsilon_{\BdKPi}}{\varepsilon_{\BPPbar}}\,
               \BF(\BdKPi) \,
               \left(\times \frac{f_d}{f_s}\right)\,,
\label{eq:BF}
\end{equation}
\noindent
where $N$ represents yields determined from fits to the \PPbar or \KPi
invariant mass distributions,
$f_d/f_s$ (included only for the $\Bs$ mode)
is the ratio of \bquark-quark hadronization probabilities into the \Bz and \Bs mesons~\cite{fsfd}
and $\varepsilon$ represents the geometrical acceptance, reconstruction and selection efficiencies.
The notation \BPPbar stands for either \BdPPbar or \BsPPbar.
The inclusion of charge-conjugate processes is implied, unless otherwise indicated.

The data sample analyzed corresponds to an integrated luminosity of 1.0\invfb
of proton-proton collision data collected by the \lhcb experiment
at center-of-mass energies of 7\tev in 2011 and 2.0\invfb at 8\tev in 2012.
The \lhcb detector~\cite{Alves:2008zz,LHCb-DP-2014-002} is a single-arm forward
spectrometer covering the \mbox{pseudorapidity} range $2<\eta <5$,
designed for the study of particles containing \bquark or \cquark
quarks. The detector includes a high-precision tracking system
consisting of a silicon-strip vertex detector surrounding the $pp$
interaction region~\cite{LHCb-DP-2014-001}, a large-area silicon-strip detector located
upstream of a dipole magnet with a bending power of about
$4{\mathrm{\,Tm}}$, and three stations of silicon-strip detectors and straw
drift tubes~\cite{LHCb-DP-2013-003} placed downstream of the magnet.
The tracking system provides a measurement of momentum, \ptot, of charged particles with
a relative uncertainty that varies from 0.5\% at low momentum to 1.0\% at 200\gevc.
The minimum distance of a track to a primary vertex (PV), the impact parameter (IP),
is measured with a resolution of $(15+29/\pt)\mum$,
where \pt is the component of the momentum transverse to the beam, in\,\gevc.
Different types of charged hadrons are distinguished using information
from two ring-imaging Cherenkov detectors~\cite{LHCb-DP-2012-003}.
Photons, electrons and hadrons are identified by a calorimeter system consisting of
scintillating-pad and preshower detectors, an electromagnetic
calorimeter and a hadronic calorimeter. Muons are identified by a
system composed of alternating layers of iron
and detector planes consisting of multiwire proportional chambers and gas electron multipliers.
Simulated data samples, produced as described
in Refs.~\cite{Sjostrand:2007gs,*Sjostrand:2006za,LHCb-PROC-2010-056,Lange:2001uf,Golonka:2005pn,Allison:2006ve,*Agostinelli:2002hh,LHCb-PROC-2011-006},
are used to evaluate the response of the detector and to investigate and characterize
possible sources of background.

Candidates are selected in a similar way for both signal \BPPbar decays
and the normalization channel \BdKPi.
Real-time event selection is performed by a
trigger~\cite{LHCb-DP-2012-004} consisting of a hardware stage,
based on information from the calorimeter and muon systems, followed
by a software stage, which performs a full event reconstruction.
The hardware trigger stage requires events to have a
hadron, photon or electron with high transverse energy (above a few \gev) deposited in the calorimeters,
or a muon with high transverse momentum.
For this analysis, the hardware trigger decision can be made either on the signal candidates
or on other particles in the event.
The software trigger requires a two-track
secondary vertex with a significant displacement from the PVs.
At least one charged particle
must have high \pt and be inconsistent with originating from a PV.
A multivariate algorithm~\cite{BBDT} is used for
the identification of secondary vertices consistent with the decay
of a \bquark or \cquark hadron.

The final selection of candidates in the signal and normalization modes
is carried out with a preselection stage, particle identification (PID) criteria,
and a requirement on the response of a multilayer perceptron (MLP) classifier~\cite{MLPs}.
To avoid potential biases, \PPbar candidates with invariant mass in the range $[ 5230, 5417 ]\mevcc$
(a $\pm 50\mevcc$ window approximately three times the invariant mass resolution around the known \Bd and
\Bs masses~\cite{PDG2016}) were not examined until the analysis procedure was finalized.

At the preselection stage, the \Bds decay products are associated with tracks with good reconstruction quality
that have $\chisqip > 9$ with respect to any PV,
where the \chisqip is defined as the difference between the vertex-fit \chisq
of a PV reconstructed with and without the track in question.
The minimum \pt of the decay products is required to be above 900\mevc
and at least one of the decay products is required to have $\pt > 2100\mevc$.
A loose PID requirement, based primarily on information from the Cherenkov detectors,
is also imposed on both particles.
The \Bds candidate must have a vertex with good reconstruction quality, $\pt > 1000 \mevc$
and a $\chisqip < 36$ with respect to the associated PV.
The associated PV is that with which it forms the smallest \chisqip. 
The angle $\theta_B$ between the momentum vector of the \Bds candidate and the line
connecting the associated PV and the candidate's decay vertex is required to be close to zero
($\cos(\theta_B) > 0.9995$).

After preselection, tight PID requirements are applied to the two final-state particles to suppress
so-called combinatorial background formed from the accidental associations of tracks unrelated to the signal decays,
and to reduce contamination from $b$-hadron decays where one or more decay products are misidentified.
The PID requirements are determined by optimizing
the figure of merit $\varepsilon^{\rm sig} / (\frac{a}{2} + \sqrt{N_{\rm bkg}})$~\cite{Punzi:2003bu},
where $a=5$ quantifies the target level of significance in standard deviations and
$\varepsilon^{\rm sig}$ is the PID efficiency of the signal selection.
The quantity $N_{\rm bkg}$ denotes the expected number of background events in the signal region.
This is estimated by extrapolating the result of a fit to the invariant mass distribution of the data
in the sideband regions above and below the signal region.
The PID criteria are allowed to be different for protons and antiprotons.
The optimization of the PID criteria applied to the normalization decay
candidates relies on maximizing the signal significance, while minimizing the contamination
from misidentified backgrounds.

Further separation between signal and combinatorial background candidates relies on
an MLP implemented with the TMVA toolkit~\cite{TMVA}.
There are ten input quantities to the MLP classifier:
the minimum values of the \pt and $\eta$ of the decay products, the scalar sum of their \pt values,
the \chisqip of the decay products;
the distance of closest approach between the two decay products;
a parameter expressing the quality of the \Bds vertex fit;
the \chisqip and $\theta_B$ angle of the \Bds candidate;
and the \pt asymmetry within a cone around the \Bds direction defined by
$A_{\pt} = (p_{\mathrm{T}}^{B} - p_{\mathrm{T}}^{\mathrm{cone}}) / (p_{\mathrm{T}}^{B} + p_{\mathrm{T}}^{\mathrm{cone}})$,
where $p_{\mathrm{T}}^{\mathrm{cone}}$ is the transverse component of the vector sum of the momenta of
all tracks measured within the cone radius $R=1.0$ around the \Bds direction,
except for the \Bds decay products.
The cone radius is defined in pseudorapidity and azimuthal angle
$( \eta, \phi)$ as $R = \sqrt{ (\Delta\eta)^{2} + (\Delta\phi)^{2} }$.
The $A_{\pt}$ requirement exploits the relative isolation of signal decay products
as compared with background.
The MLP is trained using simulated \BdPPbar decays
and data candidates in the \PPbar invariant mass sideband above 5417\mevcc
to represent the background. The requirement on the MLP response is optimized
using the same figure of merit as that used for the optimization of
the PID selection.
The MLP selection keeps approximately 60\% of the signal candidates, while
suppressing combinatorial background by two orders of magnitude.
The MLP applied to the normalization decay
candidates is the same as that trained to select the \BPPbar signal
candidates, with the requirement on the response chosen to maximize
the \BdKPi significance. 
A vanishingly small fraction of events contains a second candidate
after all selection requirements are applied and all candidates are kept.

Large control data samples of kinematically identified pions, kaons and protons
originating from the decays $\Dz \to \Km \pip$, \LPPi  and \LcPKPi
are employed to determine the efficiency of the PID requirements~\cite{LHCb-DP-2012-003}.
All the other components of the selection efficiencies are determined from simulation.
The agreement between data and simulation is verified comparing kinematic distributions
from selected \BdKPi decays.
The distributions in data are obtained with the \sPlot technique~\cite{Pivk:2004ty}
with the \Bd candidate invariant mass used as the discriminating variable.
The overall efficiencies of this analysis, including the trigger selection and the reconstruction,
are of the order $10^{-3}$.

Sources of noncombinatorial background to the \PPbar spectrum are investigated
using simulation samples. These sources include partially reconstructed backgrounds
in which one or more particles from the decay of a $b$ hadron are not associated with the
signal candidate, or $b$-hadron decays where one or more decay products are misidentified.
The sum of such backgrounds does not peak in the \Bd and \Bs signal regions
but rather contributes a smooth \PPbar mass spectrum,
which is indistinguishable from the dominant combinatorial background.

The yields of the signal and normalization candidates
are determined using unbinned maximum likelihood fits to the invariant mass distributions.
The \PPbar invariant mass distribution is described with three components, namely
the \BdPPbar and \BsPPbar signals and combinatorial background.
The \BPPbar signals are modeled with the sum of two Crystal Ball (CB)
functions~\cite{Skwarnicki:1986xj} describing the high- and low-mass
asymmetric tails.
The two components of each signal share the same peak and core width parameters.
The core widths are fixed using \BPPbar simulated samples.
A scaling factor is applied to account for differences in the resolution between data and simulation
as determined from \BdKPi candidates.
The \BsPPbar signal peak value is set relative to the \BdPPbar signal peak value determined from the fit
according to the \Bs--\Bd mass difference~\cite{PDG2016}.
The tail parameters and the relative normalization of the CB functions
are determined from simulation.
The combinatorial background is described
with a linear function, with the slope parameter allowed to vary in the fit.

\begin{figure}[tb]
  \centering
  \scalebox{.6}{\includegraphics{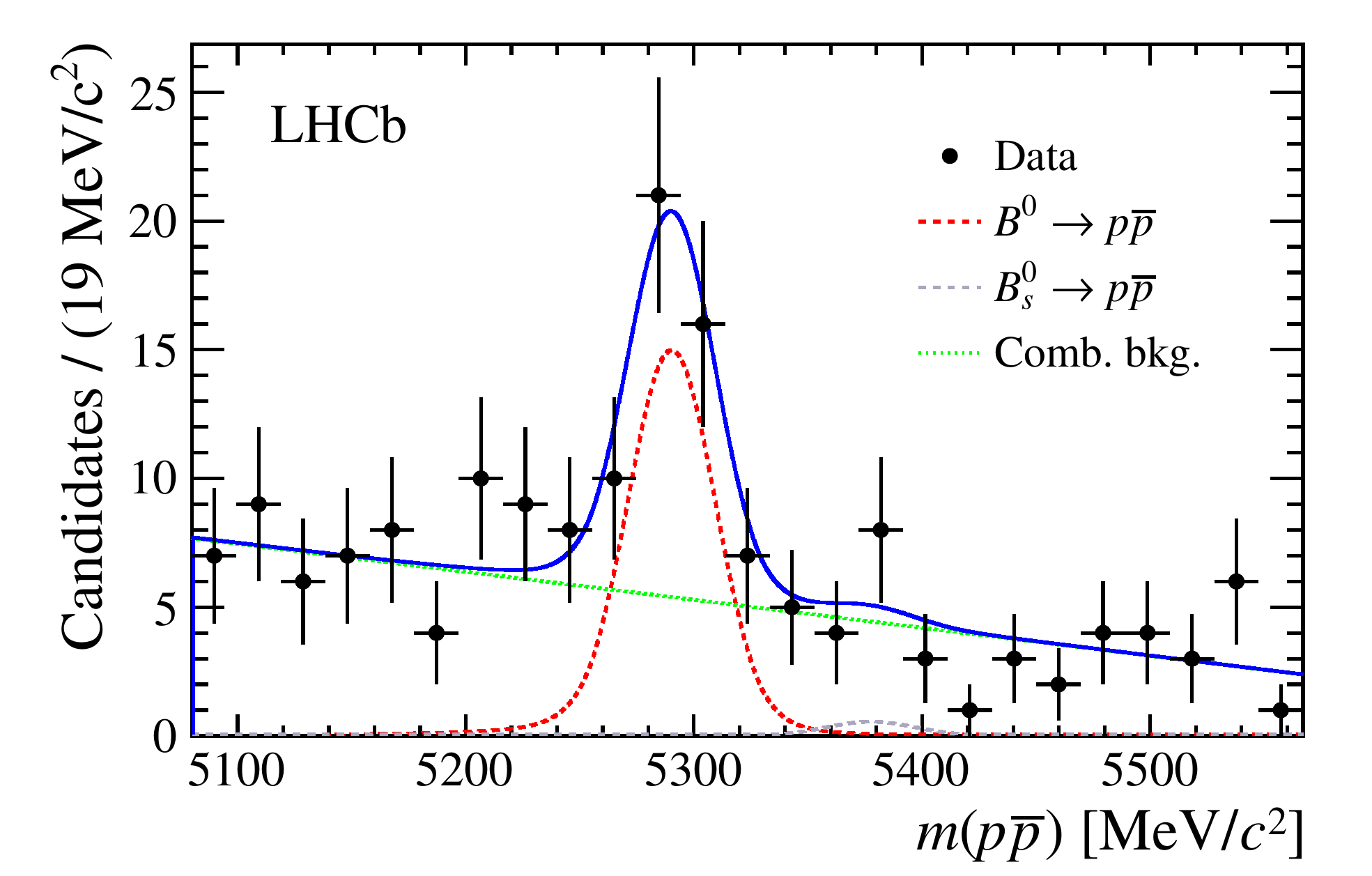}}
  \caption{Invariant mass distribution of \PPbar candidates.
           The fit result (blue, solid line) is shown together with each fit model component:
           the \BdPPbar signal (red, dashed line), the \BsPPbar signal (grey, dashed line)
           and the combinatorial background (green, dotted line).}
  \label{fig:PPbarFit}
\end{figure}

The \PPbar invariant mass distribution is presented in
Fig.~\ref{fig:PPbarFit} together with the result of the fit. The yields of the \BPPbar~signals are
$N(\BdPPbar) = 39 \pm 8$ and $N(\BsPPbar) = 2 \pm 4$,
where the uncertainties are statistical only.
The significance of each of the signals is determined from the change in
the logarithm of the likelihood between fits with and without the signal component~\cite{Wilks:1938dza}.
The \BdPPbar decay mode is found to have a significance of 5.3 standard deviations, including systematic uncertainties,
and the \BsPPbar mode is found to have a significance of 0.4 standard deviations, where, given its size,
the significance has been evaluated ignoring systematic effects.
The high significance of the \BdPPbar signal implies the first observation of a two-body charmless baryonic \Bd decay.

\begin{figure}[tbhp]
  \centering
  \scalebox{.6}{\includegraphics{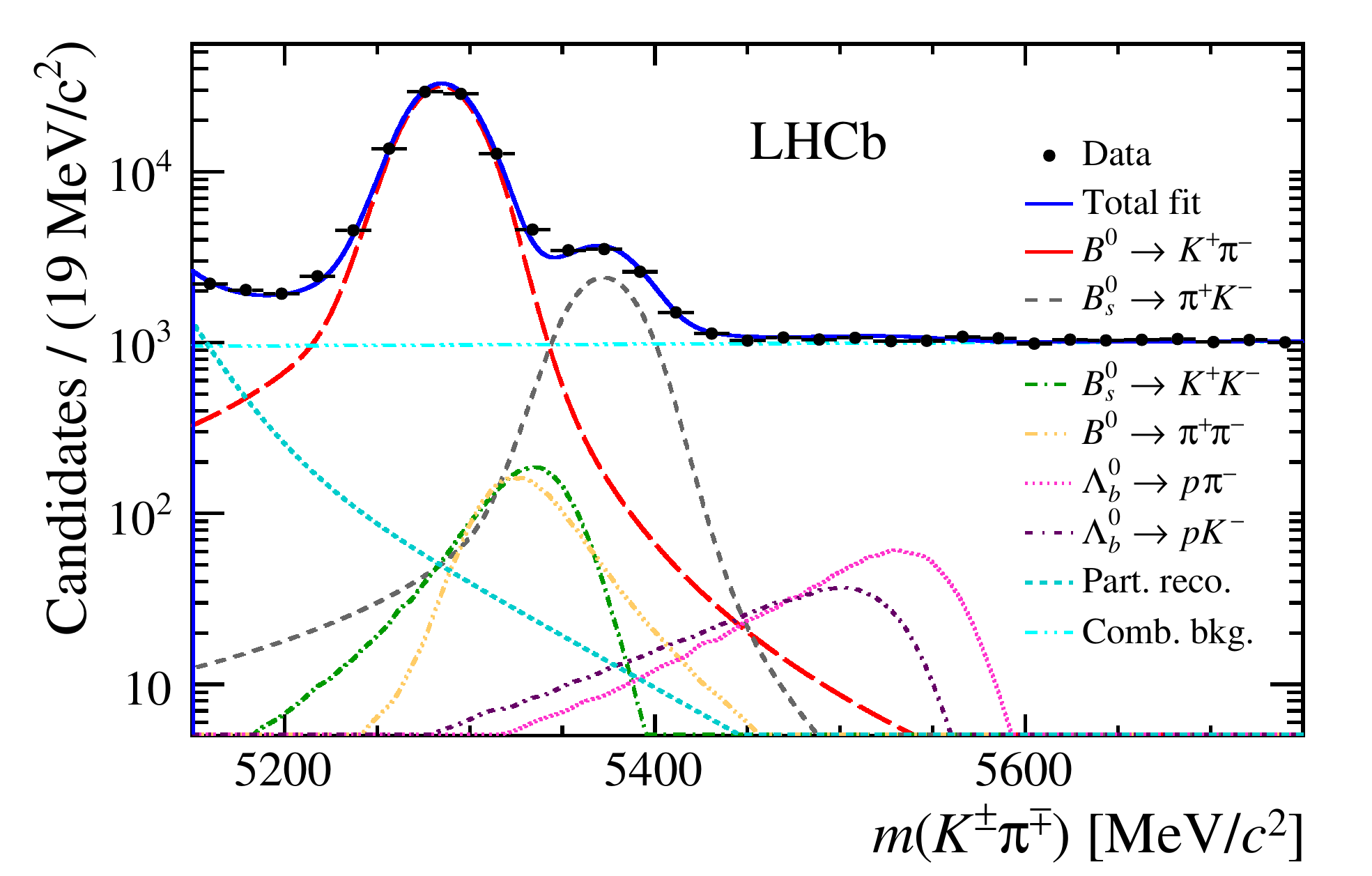}}
  \caption{Invariant mass distribution of \BothChargeKPi candidates.
           The fit result (blue, solid line) is shown together with each fit model component.}
  \label{fig:KPiFit}
\end{figure}

The \KPi invariant mass distribution of the normalization decay
candidates is described with components accounting for the \BdKPi and
\BsPiK signals; the background due to the decays \BdPiPi, \BsKK, \LbPPi and \LbPK
when at least one of the final-state particles is misidentified;
background from partially reconstructed $b$-hadron decays; and combinatorial background.
The \BdKPi and \BsPiK decays are modeled with the sum of two CB functions
sharing the same peak and core width parameters.
The peak value and core width of the \BdKPi
signal model are free parameters in the fit. The difference between the
peak positions of the \BdKPi and \BsPiK signals is constrained to its
known value~\cite{PDG2016} and the core width of the \BsPiK signal is
related to the \BdKPi signal core width by a scaling factor of 1.02 as determined from simulation.
The tail parameters and the relative normalization of
both double CB functions are determined from simulation. The
invariant mass distributions of the four misidentified decays are
determined from simulation and are modeled with nonparametric functions~\cite{Cranmer:2000du}.
The relative fractions of these background components
depend upon the branching fractions,
$b$-hadron hadronization probabilities and misidentification
rates of the backgrounds. The fractions are Gaussian-constrained to the product of these
three factors, with the widths of the Gaussian functions equal to their combined uncertainties.
The misidentification rates are determined
from calibration data samples, whereas all other selection efficiencies
are obtained from simulation. Partially reconstructed backgrounds
represent decay modes misreconstructed as signal with one or more
undetected final-state particles, possibly in conjunction with
misidentifications. The shapes of these backgrounds in \KPi invariant mass are determined from
simulation, where each contributing decay is assigned a weight
dependent on its relative branching fraction, hadronization probability
and selection efficiency. The weighted sum of the
partially reconstructed backgrounds is well modeled with the sum of
two exponential functions, the slope parameters of which are fixed from
simulation, while the yield is determined in the fit to the data.
As for the signal fit, the combinatorial background is described with a linear
function, with the slope parameter allowed to vary in the fit.

The fit to the \KPi invariant mass, shown in Fig.~\ref{fig:KPiFit}, involves seven fitted parameters
and yields \mbox{$N(\BdKPi) = 88\,961 \pm 341$} signal decays,
where the uncertainty is statistical only.

The sources of systematic uncertainty on the \BPPbar branching fractions
arise from the fit model, the limited knowledge of the selection efficiencies,
and the uncertainties on the \BdKPi branching fraction and on the ratio of
$b$-quark hadronization probabilities $f_s/f_d$.
Pseudoexperiments are used to estimate the effects of using alternative shapes
for the fit components and of including additional backgrounds in the fit.
Systematic uncertainties on the fit models are also assessed by varying
the fixed parameters of the models within their uncertainties.
The description of the combinatorial background is replaced by an exponential function.
In the fit to the signal modes, the partially reconstructed decays \BuPPbarLNu, where $\ell$ stands for an electron or a muon
and \neul for the corresponding neutrino, are added to the fit model.
Intrinsic biases in the fitted yields are also investigated with
pseudoexperiments and are found to be negligible.

Uncertainties on the efficiencies arise from
residual differences between data and simulation in the trigger, reconstruction,
selection and uncertainties on the data-driven particle identification efficiencies.
These differences are assessed using the \BdKPi normalization decay,
comparing the level of agreement between simulation and data.
The distributions of selection variables for \BdKPi signal candidates in data
are obtained by subtracting the background using the \sPlot technique~\cite{Pivk:2004ty},
with the \KPi candidate invariant mass as the discriminating variable.
The effect of binning the PID calibration samples used to obtain the PID efficiencies
is evaluated by varying the binning scheme and by adding an extra dimension accounting
for event multiplicity to the binning of the samples.

The uncertainty on the branching fraction of the normalization decay,
\mbox{$\BF(\BdKPi) = (1.96 \pm 0.05) \times 10^{-5}$~\cite{PDG2016}},
is taken as a systematic uncertainty from external inputs.
The uncertainty on the measurement $f_s/f_d = 0.259 \pm 0.015$~\cite{fsfd} is quoted as a separate source
of systematic uncertainty from external inputs in the determination of
the upper limit on $\BF(\BsPPbar)$.
The total systematic uncertainty on the \BdPPbar (\BsPPbar) branching fraction is given by the sum
of all uncertainties added in quadrature and amounts to 14.2\% (209\%).
The systematic uncertainties on the \BdPPbar (\BsPPbar) branching fraction are dominated by the uncertainties on the fit model,
which are 7.3\% (208\%), and on the reconstruction and selection efficiencies, which amount to
6.1\% (6.1\%) and 8.6\% (8.3\%), respectively.
Specifically, the systematic uncertainty arising from the description of the fit model backgrounds
dominates the uncertainty on the \BsPPbar branching fraction.

In summary, the first observation of the simplest decay of a \Bz meson to a purely baryonic
final state, \BdPPbar, is reported  using a data sample of proton-proton collisions
collected with the \lhcb experiment, corresponding to a total integrated luminosity of 3.0\invfb.
This rare two-body charmless baryonic decay is observed with a significance
of 5.3 standard deviations, including systematic uncertainties.
The \BdPPbar branching fraction is determined to be
\begin{equation}
\BF(\BdPPbar) = ( 1.25 \pm 0.27 \pm 0.18 ) \times 10^{-8} \,,\nonumber
\end{equation}
where the first uncertainty is statistical and the second systematic.
Since no \BsPPbar signal is seen, the world's best upper limit
\mbox{$\BF( \BsPPbar) < 1.5 \times 10^{-8}$} at 90\% confidence level
is set on the decay branching fraction using
the Feldman-Cousins frequentist method~\cite{PhysRevD.57.3873}.

The first observation of the decay \BdPPbar,
the rarest \Bd decay ever observed,
provides valuable input towards the understanding
of the dynamics of hadronic $B$ decays.
This measurement helps to discriminate among several QCD-based models
and makes it possible to extract both tree and penguin amplitudes of charmless two-body baryonic
decays when combining the information on the \BdPPbar and \BuPLbar branching fractions~\cite{Chua:2016aqy}.
The measured \BdPPbar branching fraction is compatible with recent theoretical calculations,
as is the upper limit on the \BsPPbar branching fraction~\cite{Hsiao:2014zza,Cheng:2014qxa,Chua:2016aqy}.
An improved measurement of the \BsPPbar branching fraction will make it possible to quantitatively
compare the models proposed in Refs.~\cite{Hsiao:2014zza} and~\cite{Chua:2016aqy}.

%
%
\section*{Acknowledgements}

\noindent We express our gratitude to our colleagues in the CERN
accelerator departments for the excellent performance of the LHC. We
thank the technical and administrative staff at the LHCb
institutes. We acknowledge support from CERN and from the national
agencies: CAPES, CNPq, FAPERJ and FINEP (Brazil); MOST and NSFC
(China); CNRS/IN2P3 (France); BMBF, DFG and MPG (Germany); INFN
(Italy); NWO (The Netherlands); MNiSW and NCN (Poland); MEN/IFA
(Romania); MinES and FASO (Russia); MinECo (Spain); SNSF and SER
(Switzerland); NASU (Ukraine); STFC (United Kingdom); NSF (USA).  We
acknowledge the computing resources that are provided by CERN, IN2P3
(France), KIT and DESY (Germany), INFN (Italy), SURF (The
Netherlands), PIC (Spain), GridPP (United Kingdom), RRCKI and Yandex
LLC (Russia), CSCS (Switzerland), IFIN-HH (Romania), CBPF (Brazil),
PL-GRID (Poland) and OSC (USA). We are indebted to the communities
behind the multiple open-source software packages on which we depend.
Individual groups or members have received support from AvH Foundation
(Germany), EPLANET, Marie Sk\l{}odowska-Curie Actions and ERC
(European Union), ANR, Labex P2IO, ENIGMASS and OCEVU, and R\'{e}gion
Auvergne-Rh\^{o}ne-Alpes (France), RFBR and Yandex LLC (Russia), GVA,
XuntaGal and GENCAT (Spain), Herchel Smith Fund, the Royal Society,
the English-Speaking Union and the Leverhulme Trust (United Kingdom).

\addcontentsline{toc}{section}{References}
\setboolean{inbibliography}{true}
\bibliographystyle{LHCb}
\bibliography{main,LHCb-PAPER,LHCb-CONF,LHCb-DP,LHCb-TDR}
 

\clearpage
\centerline{\large\bf LHCb collaboration}
\begin{flushleft}
\small
R.~Aaij$^{40}$,
B.~Adeva$^{39}$,
M.~Adinolfi$^{48}$,
Z.~Ajaltouni$^{5}$,
S.~Akar$^{59}$,
J.~Albrecht$^{10}$,
F.~Alessio$^{40}$,
M.~Alexander$^{53}$,
A.~Alfonso~Albero$^{38}$,
S.~Ali$^{43}$
% ...

{\footnotesize \it
% ...
$ ^{5}$Clermont Universit{\'e}, Universit{\'e} Blaise Pascal, CNRS/IN2P3, LPC, Clermont-Ferrand, France\\
% ...
$ ^{10}$Fakult{\"a}t Physik, Technische Universit{\"a}t Dortmund, Dortmund, Germany\\
% ...
$ ^{38}$ICCUB, Universitat de Barcelona, Barcelona, Spain\\
$ ^{39}$Universidad de Santiago de Compostela, Santiago de Compostela, Spain\\
$ ^{40}$European Organization for Nuclear Research (CERN), Geneva, Switzerland\\
% ...
$ ^{43}$Nikhef National Institute for Subatomic Physics, Amsterdam, The Netherlands\\
% ...
$ ^{48}$H.H. Wills Physics Laboratory, University of Bristol, Bristol, United Kingdom\\
% ...
$ ^{53}$School of Physics and Astronomy, University of Glasgow, Glasgow, United Kingdom\\
% ...
$ ^{59}$University of Cincinnati, Cincinnati, OH, United States\\
% ...
}
\end{flushleft}

\end{document}